\documentclass[prd,aps,eqsecnum,12pt,onecolumn]{revtex4-2}

\usepackage{hyperref,amsfonts,amsmath,amssymb,bm,graphicx}

\begin{document}

\author{Maxim Dvornikov}
\email{maxim.dvornikov@gmail.com}

\title{Quantum field theory treatment of the spin-flavor precession of Dirac
neutrinos with diagonal magnetic moments}

\affiliation{Pushkov Institute of Terrestrial Magnetism, Ionosphere
and Radiowave Propagation (IZMIRAN),
108840 Moscow, Troitsk, Russia}

%
%
%
%
%
%
%
%

\begin{abstract}
We apply the quantum field theory (QFT) approach to describe the spin-flavor
precession of Dirac neutrinos. The magnetic moments matrix of these
neutrinos is taken to be diagonal in the mass basis. Neutrino mass
eigenstates are considered to be virtual particles in the QFT approach.
The neutrino contribution to the matrix element of the overall process,
which includes the emission, the propagation and the detection of
neutrinos, is contained in the propagators of massive neutrinos. Using
the recently derived propagator of a massive neutrino, which exactly
accounts for the magnetic moment interaction, we derive the transition
probability of the process $\nu_{e\mathrm{L}}\to\nu_{\mu\mathrm{L}}$
under the influence of a magnetic field. The result is compared with
the prediction of the standard quantum mechanical (QM) approach. We
briefly discuss the possibility to derive the QFT corrections to the
QM expressions.
\end{abstract}

\maketitle


\section{Introduction}

The flavor content in a neutrino beam was established experimentally
to be the subject of a periodic change while a beam propagates in
space. This process is called neutrino flavor oscillations. Besides
flavor oscillations, polarization of neutrinos can be altered in the
presence of an external electromagnetic field. It is called the neutrino
spin-flavor precession. Originally, it was proposed in Refs.~\cite{LimMar88,Akh88}.

The interaction of neutrinos with a magnetic field is owing to the
nonzero magnetic moments. The properties of the magnetic
moments matrix are different for Dirac and Majorana particles. Unlike
Majorana neutrinos, Dirac particles, which are considered in the present
work, can have diagonal magnetic moments. We also mention that diagonal
magnetic moments are not suppressed by the Glashow--Iliopoulos--Maiani
mechanism (see, e.g., Ref.~\cite{FukYan03}). The recent review on
the neutrino electrodynamics is provided in Ref.~\cite{Giu25}.

As a rule, neutrino oscillations are studied with the quantum mechanical
(QM) approach. However, this kind of treatment of neutrino oscillations
has certain shortcomings which are summarized in Ref.~\cite{NauNau20}.
A quantum field theory
(QFT) description should be developed to overcome the difficulties of the QM approach. The QFT approach was originally
proposed in Ref.~\cite{Kob82} Then, it was independently rediscovered
in Refs.~\cite{GiuKimLee93,GriSto96} In this approach, neutrino
mass eigenstates are considered to be virtual particles described
by their propagators.

The extension of the approach in Refs.~\cite{Kob82,GiuKimLee93,GriSto96}
to cover the situations of the neutrino interaction with external
fields was developed in Refs.~\cite{Dvo25,Dvo26b,Dvo26a}. In particular,
we are interested in Refs.~\cite{Dvo26b,Dvo26a} where the spin-flavor
precession of Majorana and Dirac neutrinos in a magnetic field were
considered. Note that the situations studied in Refs.~\cite{Dvo26b,Dvo26a}
imply significant transition magnetic moments.

The study of the spin-flavor precession of Dirac neutrinos with arbitrary
magnetic moments in frames of the relativistic QM was made in Ref.~\cite{DvoMaa07}.
Some results of Ref.~\cite{DvoMaa07} were also rediscovered in Ref.~\cite{PopStu19}.
The spin-flavor
precession of Dirac neutrinos with diagonal magnetic moments was studied
in Ref.~\cite{EgoVol22} in frames of the QFT. However, unlike Ref.~\cite{EgoVol22},
we do not consider the modification of neutrino propagators (cf. Refs.~\cite{Dvo25,Dvo26b,Dvo26a})
which are constructed in the standard manner.

In the present work, we continue the study of the neutrino spin-flavor
precession in frames of the QFT, which was started in Refs.~\cite{Dvo25,Dvo26b,Dvo26a}.
In Sec.~\ref{sec:QFT}, we briefly remind the basics of the neutrino
masses and mixing, as well as of the electrodynamics of Dirac neutrinos.
Then, we apply the QFT based approach to the neutrino spin-flavor
precession. Using the recently obtained propagator of a massive Dirac
neutrino with a diagonal magnetic moment, we derive the matrix element
and the transition probability for one of the channels of the spin-flavor
precession. We compare our findings with the results of the standard
QM description of the spin-flavor precession in Sec.~\ref{sec:QM}.
Finally, we conclude in Sec.~\ref{sec:CONCL}.

\section{QFT description of the spin-flavor precession}\label{sec:QFT}

In this section, we apply QFT to study the spin-flavor precession
of Dirac neutrinos having only the diagonal magnetic moments.

Flavor, $\nu_{\lambda}$, and mass, $\psi_{a}$, eigenstates of neutrinos
are related by
\begin{equation}
\nu_{\lambda}=\sum_{a=1,2}U_{\lambda a}\psi_{a},\quad\lambda=e,\mu,\quad(U_{\lambda a})=\left(\begin{array}{cc}
\cos\theta & \sin\theta\\
-\sin\theta & \cos\theta
\end{array}\right),\label{eq:nupsi}
\end{equation}
where $\theta$ is the vacuum mixing angle. In Eq.~(\ref{eq:nupsi}),
we assume that $\psi_{a}$ are Dirac particles. Moreover, we deal with two types of neutrinos in Eq.~(\ref{eq:nupsi}). The mixing matrix $(U_{\lambda a})$
in Eq.~(\ref{eq:nupsi}) is chosen in such a way to diagonalize the
mass matrix of flavor neutrinos.

We suppose that the neutrino mass eigenstates possess the magnetic
moments $\mu_{a}$. The Lagrangian for $\psi_{a}$ in the presence
of the electromagnetic field $F_{\mu\nu}=(\mathbf{E},\mathbf{B})$
has the form,
\begin{equation}
\mathcal{L}=\sum_{a}\bar{\psi}_{a}\left[\mathrm{i}\gamma^{\mu}\partial_{\mu}-m_{a}-\frac{\mu_{a}}{2}F_{\mu\nu}\sigma^{\mu\nu}\right]\psi_{a},\label{eq:Lagrpsi}
\end{equation}
where $\gamma^{\mu}=(\gamma^{0},\bm{\gamma})$ and $\sigma_{\mu\nu}=\tfrac{\mathrm{i}}{2}[\gamma_{\mu},\gamma_{\nu}]_{-}$
are the Dirac matrices, and $m_{a}$ is the mass of $\psi_{a}$.

\begin{figure}
  \centering
  \includegraphics{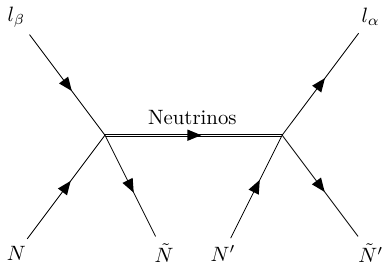}
  \protect
\caption{The Feynman diagram corresponding to the $S$-matrix element in Eq.~(\ref{eq:Smatr}).
Virtual neutrinos are depicted by a broad line since their propagator
accounts for the contribution of a magnetic field.}\label{fig:feyndiag}
\end{figure}

The QFT approach to neutrino oscillations involves both the production,
and the propagation, and the detection of neutrinos. We assume that
a source and a detector of neutrinos consists of heavy nuclei $N$
and $N'$ which become $\tilde{N}$ and $\tilde{N}'$ as a result of the interaction
with charged leptons $l_{\beta}$ and $l_{\alpha}$. The marker of
neutrino oscillations is when $l_{\alpha}\neq l_{\beta}$. One can
see that neutrinos are virtual particles in this approach. The overall
process is shown in Fig~\ref{fig:feyndiag}.

Therefore, one should calculate the following $S$-matrix element:
\begin{equation}
S=-\frac{1}{2}\left(\sqrt{2}G_{\mathrm{int}}\right)^{2}\int\mathrm{d}^{4}x\mathrm{d}^{4}y\left\langle \tilde{N},\tilde{N}',l_{\alpha}\left|T\left\{ j_{\mu}^{\dagger}(x)J^{\mu}(x)j^{\nu}(y)J_{\nu}^{\dagger}(y)\right\} \right|N,N',l_{\beta}\right\rangle ,\label{eq:Smatr}
\end{equation}
where
\begin{equation}
j_{\mu}=\sum_{\lambda}\bar{\nu}_{\lambda\mathrm{L}}\gamma_{\mu}l_{\lambda\mathrm{L}},\label{eq:lepcurr}
\end{equation}
is the leptonic current, $J^{\mu}$ is the nuclear current, and $G_{\mathrm{int}}$
is the coupling constant. In Eq.~(\ref{eq:lepcurr}), we assume that
the emission and the detection of neutrinos is treated in frames of
the standard model. Thus, only left-handed chiral projections of leptons
are involved, e.g., $l_{\lambda\mathrm{L}}=\tfrac{1}{2}(1-\gamma^{5})l_{\lambda}$,
where $\gamma^{5}=\mathrm{i}\gamma^{0}\gamma^{1}\gamma^{2}\gamma^{3}$.

Let us study a particular process, $e\to\mu$. We take that the nuclei
in a source and a detector are quite heavy, so that they do not change
their positions $\mathbf{x}_{1}$ and $\mathbf{x}_{2}$. Then, assuming
that the incoming and outgoing leptons are the plane waves with the
four momenta $p_{e,\mu}^{\mu}=(E_{e,\mu},\mathbf{p}_{e,\mu})$ and
the bispinors $u_{e,\mu}$, we transform the $S$-matrix in Eq.~(\ref{eq:Smatr})
to the form,
\begin{equation}
S=-2\pi G_{\mathrm{int}}^{2}\delta(E_{e}-E_{\mu})e^{-\mathrm{i}\mathbf{p}_{\mu}\mathbf{x}_{2}+\mathrm{i}\mathbf{p}_{e}\mathbf{x}_{1}}\mathrm{i}\mathcal{M}_{e\to\mu},
\end{equation}
where the matrix element reads
\begin{equation}
\mathcal{M}_{e\to\mu}=\bar{u}_{\mu}\gamma_{\mathrm{L}}^{0}\sum_{a=1,2}\left[\int\frac{\mathrm{d}^{3}q}{(2\pi)^{3}}e^{\mathrm{i}\mathbf{qL}}U_{\mu a}U_{ea}^{*}S_{a}(E,\mathbf{q})\right]\gamma_{\mathrm{L}}^{0}u_{e}.\label{eq:matrelgen}
\end{equation}
Here, $E=(E_{e}+E_{\mu})/2$ is the mean energy of incoming and outgoing
leptons, $\mathbf{L}=\mathbf{x}_{2}-\mathbf{x}_{1}$ is the vector
from a source to a detector, and $\gamma_{\mathrm{L}}^{0}=\gamma_{0}(1-\gamma^{5})/2$.

Equation~(\ref{eq:Lagrpsi}) implies that the neutrino mass eigenstates
do not mix. Hence, in Eq.~(\ref{eq:matrelgen}), we have the 4D Fourier
transform $S_{a}(q^{0},\mathbf{q})$ of the diagonal propagator of
$\psi_{a}$, which is defined in the standard manner
\begin{equation}
S_{a}(x-y)=-\mathrm{i}\left\langle 0\left|T\left\{ \psi_{a}(x)\bar{\psi}_{a}(y)\right\} \right|0\right\rangle .\label{eq:proppsiadef}
\end{equation}
Since we consider the spin-flavor precession in the electromagnetic
field, the neutrino wave functions in Eq.~(\ref{eq:proppsiadef})
should exactly account for the contribution of the electromagnetic
field based on Eq.~(\ref{eq:Lagrpsi}). That is why, in Fig.~\ref{fig:feyndiag},
we depict virtual neutrinos as a broad line.

Now, we suppose that only a magnetic field, which is along the $z$-axis,
is present: $\mathbf{B}=B\mathbf{e}_{z}$ and $\mathbf{E}=0$. The
propagator of a neutrino mass eigenstate, interacting with such a
magnetic field, was found in Ref.~\cite{Dvo26d},
\begin{equation}
S_{a}(p_{0},\mathbf{p})=\frac{1}{8}\sum_{\zeta=\pm}\left(\begin{array}{cccc}
z_{+}^{2}A_{+}; & -2\Delta\frac{\zeta z_{-}z_{+}p_{3}}{E_{\zeta}}e^{-\mathrm{i}\phi}; & -2\Delta\frac{z_{+}^{2}p_{3}}{E_{a\zeta}}; & -\zeta z_{-}z_{+}A_{+}e^{-\mathrm{i}\phi}\\
-2\Delta\frac{\zeta z_{-}z_{+}p_{3}}{E_{a\zeta}}e^{\mathrm{i}\phi}; & z_{-}^{2}A_{-}; & \zeta z_{-}z_{+}A_{-}e^{\mathrm{i}\phi}; & 2\Delta\frac{z_{-}^{2}p_{3}}{E_{a\zeta}}\\
2\Delta\frac{z_{+}^{2}p_{3}}{E_{a\zeta}}; & -\zeta z_{-}z_{+}A_{-}e^{-\mathrm{i}\phi}; & -z_{+}^{2}A_{-}; & -2\Delta\frac{\zeta z_{-}z_{+}p_{3}}{E_{a\zeta}}e^{-\mathrm{i}\phi}\\
\zeta z_{-}z_{+}A_{+}e^{\mathrm{i}\phi}; & -2\Delta\frac{z_{-}^{2}p_{3}}{E_{a\zeta}}; & -2\Delta\frac{\zeta z_{-}z_{+}p_{3}}{E_{a\zeta}}e^{\mathrm{i}\phi}; & -z_{-}^{2}A_{+}
\end{array}\right).\label{eq:proppsiB}
\end{equation}
Here
\begin{align}
A_{\pm}= & \frac{(f_{+}\pm\zeta f_{-})^{2}}{p_{0}-E_{a\zeta}+\mathrm{i}0}+\frac{(f_{+}\mp\zeta f_{-})^{2}}{p_{0}+E_{a\zeta}-\mathrm{i}0},\quad\Delta=\frac{1}{p_{0}-E_{a\zeta}+\mathrm{i}0}-\frac{1}{p_{0}+E_{a\zeta}-\mathrm{i}0},\nonumber \\
z_{\pm}= & \sqrt{1\pm\zeta\frac{m_{a}}{S_{a}}},\quad f_{\pm}=\sqrt{1\pm\frac{p_{3}}{E_{a\zeta}}},\quad S_{a}=\sqrt{m_{a}^{2}+p_{1}^{2}+p_{2}^{2}},\quad\tan\phi=p_{2}/p_{1},\label{eq:propingr}
\end{align}
where $\mathrm{i}0$ means an infinitesimal imaginary term and $\zeta=\pm1$
is the sign factor in the eigenvalue of the operator, 
\begin{equation}
\Pi=m_{a}\gamma^{0}-\mathrm{i}\gamma^{0}\gamma^{5}(\bm{\Sigma}\times\mathbf{p})_{3}-\mu_{a}B,\quad\Pi\psi_{a}=(\zeta S_{a}-\mu_{a}B)\psi_{a}.\label{eq:Pioper}
\end{equation}
In Eq.~(\ref{eq:Pioper}), $\bm{\Sigma}=\gamma^{5}\gamma^{0}\bm{\gamma}$.
We take that the Dirac matrices are in the Dirac representation in
Eq.~(\ref{eq:proppsiB}). In Eqs.~(\ref{eq:proppsiB}) and~(\ref{eq:propingr}),
\begin{equation}
E_{a\zeta}=\sqrt{(S_{a}-\zeta\mu_{a}B)^{2}+p_{3}^{2}},
\end{equation}
are the energy levels of a neutrino with the mass $m_{a}$ and the
magnetic moment $\mu_{a}$ in a magnetic field. Note that the solution
of the Dirac equation, which the propagator in Eq.~(\ref{eq:proppsiB})
is based on, was found in Ref.~\cite{TerBagKha65} originally.

We should make an important comment on the observability of the spin-flavor
precession. A nonzero magnetic field causes the spin-flip of a neutrino.
Since we study Dirac neutrinos, we should deal with a right-polarized
particle in a detector. However, we assume that a detector is based
on the standard model physics. Thus, it is practically insensitive
to this kind of neutrinos. Thus, we should look for the impact of
the magnetic field, e.g., to the channel $\nu_{e\mathrm{L}}\to\nu_{\mu\mathrm{L}}$.
For this purpose, it is enough to study the matrix element in Eq.~(\ref{eq:matrelgen}).

We consider the spin-flavor precession in the transverse magnetic
field. In this case, we take that $\mathbf{L}=L\mathbf{e}_{x}$. Moreover,
we assume that both charged leptons, $e$ and $\mu$, are ultrarelativistic,
left-handed and propagate along the $x$-axis. In the Dirac representation,
they have the following properly normalized bispinors:
\begin{equation}
u_{e,\mu}=\frac{1}{2}(1,-1,1,-1)^{\mathrm{T}}.\label{eq:uemu}
\end{equation}
One can check by the direct calculation that $\gamma^{5}u_{e,\mu}=-u_{e,\mu}$
and $\Sigma_{x}u_{e,\mu}=-u_{e,\mu}$.

Using Eqs.~(\ref{eq:nupsi}), (\ref{eq:matrelgen}), and~(\ref{eq:uemu}),
one gets that
\begin{align}
\mathcal{M}_{e\to\mu}= & \frac{1}{8}\sin2\theta[(\tilde{S}_{11}^{(2)}+\tilde{S}_{22}^{(2)}-\tilde{S}_{12}^{(2)}-\tilde{S}_{21}^{(2)}-\tilde{S}_{33}^{(2)}-\tilde{S}_{44}^{(2)}+\tilde{S}_{34}^{(2)}+\tilde{S}_{43}^{(2)}\nonumber \\
 & +\tilde{S}_{31}^{(2)}+\tilde{S}_{42}^{(2)}-\tilde{S}_{32}^{(2)}-\tilde{S}_{41}^{(2)}-\tilde{S}_{13}^{(2)}-\tilde{S}_{24}^{(2)}+\tilde{S}_{14}^{(2)}+\tilde{S}_{23}^{(2)})-(2\to1)],\label{eq:matrellongsum}
\end{align}
where 
\begin{equation}
\tilde{S}_{ij}^{(a)}=\int\frac{\mathrm{d}^{3}q}{(2\pi)^{3}}e^{\mathrm{i}\mathbf{qL}}S_{ij}^{(a)}(E,\mathbf{q}),\quad i,j=1,\dots4,\label{eq:FourierSij}
\end{equation}
and $S_{ij}^{(a)}$ are the components of the matrix in Eq.~(\ref{eq:proppsiB}).
In Eq.~(\ref{eq:matrellongsum}), the term $(2\to1)$ means that
one should replace the upper index in the beforehand expression.

We choose the cylindrical coordinates $\mathbf{q}=\rho\mathbf{e}_{\rho}+z\mathbf{e}_{z}$
to calculate the integrals in Eq.~(\ref{eq:FourierSij}). Using Eq.~(\ref{eq:proppsiB}),
we rewrite the matrix element in Eq.~(\ref{eq:matrellongsum}) in
the form, 
\begin{equation}\label{eq:matrelI}
  \mathcal{M}_{e\to\mu}=I_{2,\zeta=+}+I_{2,\zeta=-}-I_{1,\zeta=+}-I_{1,\zeta=1},
\end{equation}
where
\begin{align}\label{eq:Iazeta}
I_{a\zeta}= & \frac{\sin2\theta}{8}\int\frac{\rho\mathrm{d}\rho\mathrm{d}z\mathrm{d}\phi}{(2\pi)^{3}}e^{\mathrm{i}\rho L\cos\phi}\bigg\{\left(\frac{1}{E-E_{a\zeta}+\mathrm{i}0}+\frac{1}{E+E_{a\zeta}-\mathrm{i}0}\right)\nonumber \\
 & +\frac{1}{S_{a}E_{a\zeta}}\left[\zeta m_{2}z-\rho\cos\phi(S_{a}-\zeta\mu_{a}B)\right]\left(\frac{1}{E-E_{a\zeta}+\mathrm{i}0}-\frac{1}{E+E_{a\zeta}-\mathrm{i}0}\right)\bigg\}.
\end{align}

Let us compute, e.g., the integral $I_{2+}$ in Eq.~(\ref{eq:Iazeta}).
The rest rest of the integrals in Eq.~(\ref{eq:matrelI}) is computed
analogously. Since we assume that neutrinos are ultrarelativistic,
we can neglect the antiparticle degrees of freedom in Eq.~(\ref{eq:Iazeta}).
Thus, we drop the terms $\propto\frac{1}{E+E_{a\zeta}-\mathrm{i}0}$
in the integrand. The justification of the procedure is given in Ref.~\cite{Dvo26c}.

The integrals over $\phi$ are computed with help of the known expressions,
\begin{equation}
\int_{0}^{2\pi}\frac{\mathrm{d}\phi}{2\pi}e^{\mathrm{i}\rho L\cos\phi}=J_{0}(\rho L),\quad\int_{0}^{2\pi}\frac{\mathrm{d}\phi}{2\pi}e^{\mathrm{i}\rho L\cos\phi}=\mathrm{i}J_{1}(\rho L),\label{eq:Besseldef}
\end{equation}
where $J_{0}(x)$ and $J_{1}(x)$ are the Bessel functions.

To compute the integral over $z$ we expand the denominator near the
pole $z_{0}$, $E-E_{2+}+\mathrm{i}0=-z_{0}(z-z_{0})/E+\dotsc$, where $z_{0} = \sqrt{E^{2}-\rho_{0}^{2}}+\mathrm{i}0$ if $\rho_{0}<E$, and $z_{0} = \mathrm{i}\sqrt{\rho_{0}^{2}-E^{2}}$ if $\rho_{0}>E$. 
Here $\rho_{0}=\rho_{0}(\rho)=\sqrt{\rho^{2}+m_{2}^{2}}-\mu_{2}B$.
Closing the contour in the upper half-plane and accounting for Eq.~(\ref{eq:Besseldef}), 
we rewrite $I_{2+}$ as
\begin{align}\label{eq:I2+rho}
  I_{2+}= & -\frac{\mathrm{i}E\sin2\theta}{16\pi}
  \bigg\{
    \int_{0}^{\tilde{\rho}_{0}}\frac{\rho\mathrm{d}\rho}{\sqrt{E^{2}-\rho_{0}^{2}}}
    \left[
      \left(
        1+\frac{m_{2}\sqrt{E^{2}-\rho_{0}^{2}}}{S_{2}E}
      \right)
      J_{0}(\rho L)-\mathrm{i}J_{1}(\rho L)\frac{\rho(S_{2}-\mu_{2}B)}{S_{2}E}
    \right],
    \nonumber
    \\
    & -
    \int_{\tilde{\rho}_{0}}^{+\infty}\frac{\mathrm{i}\rho\mathrm{d}\rho}{\sqrt{\rho_{0}^{2}-E^{2}}}
    \left[
      \left(
        1+\frac{\mathrm{i}m_{2}\sqrt{\rho_{0}^{2}-E^{2}}}{S_{2}E}
      \right)J_{0}(\rho L)-\mathrm{i}J_{1}(\rho L)\frac{\rho(S_{2}-\mu_{2}B)}{S_{2}E}
    \right]
  \bigg\},
\end{align}
where $\tilde{\rho}_{0}$ is defined by $E=\sqrt{\tilde{\rho}_{0}^{2}+m_{a}^{2}}-\mu_{2}B$.

If we assume that neutrinos are ultrarelativistic with $m_{a}\ll E$
and the magnetic interaction is rather weak, $\mu_{a}B\ll E$, one
gets that 
\begin{equation}
\tilde{\rho}_{0}\approx E+\mu_{2}B-\frac{m_{2}^{2}}{2E},\label{eq:tilderho}
\end{equation}
Now, let us consider one of the integrals in Eq.~(\ref{eq:I2+rho}),
namely
\begin{equation}
\mathcal{I}=\int_{0}^{\tilde{\rho}_{0}}\frac{\rho\mathrm{d}\rho J_{0}(\rho L)}{\sqrt{E^{2}-\rho_{0}^{2}}}.\label{eq:calI}
\end{equation}
In case of ultrarelativistic neutrinos interacting with a weak magnetic
field, we can represent $\mathcal{I}$ in Eq.~(\ref{eq:calI}) as
$\mathcal{I}\approx\mathcal{I}_{0}+\mathcal{I}_{1}(EL)$, where 
\begin{align}
\mathcal{I}_{0} & =\int_{0}^{\tilde{\rho}_{0}}\frac{\rho J_{0}(\rho L)\mathrm{d}\rho}{\sqrt{\tilde{\rho}_{0}^{2}-\rho^{2}}}=\frac{\sin(\tilde{\rho}_{0}L)}{L},\nonumber \\
\mathcal{I}_{1} & =\frac{1}{L}\frac{\mu_{2}B}{E}f_{1}(y),\quad f_{1}(y)=y\int_{0}^{y}\frac{xJ_{0}(x)\mathrm{d}x}{(y+x)\sqrt{y^{2}-x^{2}}},\label{eq:I0I1}
\end{align}
where $\tilde{\rho}_{0}$ is given in Eq.~(\ref{eq:tilderho}) and
$y=EL$. As a rule, one studies neutrino oscillations at a large distance
$L\gg E^{-1}$. In this situation, one can show that, at $y\gg1$,
$|f_{1}|<1$ in Eq.~(\ref{eq:I0I1}). Therefore, $\mathcal{I}_{1}$
is a small quantity compared to $\mathcal{I}_{0}$.

Computing the rest of the integrals in Eq.~(\ref{eq:I2+rho}) in
the same manner as we made for $\mathcal{I}$ in Eq.~(\ref{eq:calI}),
as well as using facts that
\begin{equation}
\int_{\tilde{\rho}_{0}}^{\infty}\frac{\rho J_{0}(\rho L)\mathrm{d}\rho}{\sqrt{\rho^{2}-\tilde{\rho}_{0}^{2}}}=\frac{\cos(aL)}{L},
\end{equation}
and $J_{0}'(x)=-J_{1}(x)$, one gets that
\begin{equation}
I_{2+}\approx-\frac{E\sin2\theta}{8\pi L}\exp\left[\mathrm{i}\left(E+\mu_{2}B-\frac{m_{2}^{2}}{2E}\right)L\right].
\end{equation}
The remaining integrals in Eq.~(\ref{eq:matrelI}) can be calculated
similarly.

We provide the final expression for the matrix element as
\begin{align}\label{eq:matrelfin}
  \mathcal{M}_{e\to\mu}= & -\frac{Ee^{-\mathrm{i}EL}}{2\pi L}\frac{\sin2\theta}{2}
  \bigg[
    \cos(\mu_{2}BL)\exp\left(-\mathrm{i}\frac{m_{2}^{2}}{2E}L\right)
    \notag
    \\
    & -
    \cos(\mu_{1}BL)\exp\left(-\mathrm{i}\frac{m_{1}^{2}}{2E}L\right)
  \bigg].
\end{align}
The transition probability for the channel $\nu_{e\mathrm{L}}\to\nu_{\mu\mathrm{L}}$
of the spin-flavor precession is $P_{\nu_{e\mathrm{L}}\to\nu_{\mu\mathrm{L}}}\propto|\mathcal{M}_{e\to\mu}|^{2}$.
Using Eq.~(\ref{eq:matrelfin}), one gets that
\begin{align}\label{eq:PnueLmuL}
  P_{\nu_{e\mathrm{L}}\to\nu_{\mu\mathrm{L}}}(L) = & \frac{\sin^{2}(2\theta)}{4}
  \bigg[
    \cos^{2}(\mu_{1}BL)+\cos^{2}(\mu_{2}BL)
    \notag
    \\
    & -
    2\cos(\mu_{1}BL)\cos(\mu_{2}BL)\cos\left(\frac{\Delta m^{2}}{2E_{\nu}}L\right)
  \bigg],
\end{align}
where $\Delta m^{2}=m_{2}^{2}-m_{1}^{2}$. Shortly in Sec.\ref{sec:QM},
we compare the transition probability in Eq.~(\ref{eq:PnueLmuL}),
derived in frames of QFT, with the prediction of the QM treatment
of the spin-flavor precession.

\section{QM treatment of the spin-flavor precession}\label{sec:QM}

In this section, we derive the probabilities of the spin-flavor precession
within the QM and compare them with the results of Sec.~\ref{sec:QFT}.

The description of the spin-flavor precession of Dirac neutrinos with
diagonal magnetic moments $\mu_{1,2}$ is made in the mass basis.
We take the effective wave function in the form, $\Psi=(\psi_{1\mathrm{L}},\psi_{2\mathrm{L}},\psi_{1\mathrm{R}},\psi_{2\mathrm{R}})^{\mathrm{T}}$.
The effective Schrodinger equation in mass basis under the influence
of the magnetic field $B$, which is transverse to the neutrino momentum,
reads
\begin{equation}
\mathrm{i}\frac{\mathrm{d}\Psi}{\mathrm{d}t}=H\Psi,\quad H=\left(\begin{array}{cccc}
E_{1} & 0 & -\mu_{1}B & 0\\
0 & E_{2} & 0 & -\mu_{2}B\\
-\mu_{1}B & 0 & E_{1} & 0\\
0 & -\mu_{2}B & 0 & E_{2}
\end{array}\right),\label{eq:SchreqPsi}
\end{equation}
where $E_{1,2}=\sqrt{p^{2}+m_{1,2}^{2}}$, $p$ is the momentum of
neutrinos in a beam, and $m_{1,2}$ are the masses of mass eigenstates.
We assume that one has only left-polarized electron neutrinos initially.
In this situation, Eq.~(\ref{eq:SchreqPsi})
is supplied with the following initial condition: $\Psi(0)=(\cos\theta,\sin\theta,0,0)^{\mathrm{T}}$, where we account for Eq.~(\ref{eq:nupsi})

Using the results of Ref.~\cite{DvoMaa07}, where the evolution of
Dirac neutrinos with arbitrary magnetic moments was studied, one obtains
the solution of Eq.~(\ref{eq:SchreqPsi}). Based on this finding,
as well as using Eq.~(\ref{eq:nupsi}), we derive the effective wave
function in the flavor basis $N=(\nu_{e\mathrm{L}},\nu_{\mu\mathrm{L}},\nu_{e\mathrm{R}},\nu_{\mu\mathrm{R}})^{\mathrm{T}}$.
For the sake of brevity we provide just the expression for $\nu_{\mu\mathrm{L}}$,
\begin{equation}\label{eq:Nut}
  \nu_{\mu\mathrm{L}} = \sin\theta\cos\theta[e^{-\mathrm{i}E_{2}t}\cos(\mu_{2}Bt)-e^{-\mathrm{i}E_{1}t}\cos(\mu_{1}Bt)]
\end{equation}
%
The probability of the transition $\nu_{e\mathrm{L}}\to\nu_{\mu\mathrm{L}}$,
which was studied in Sec.~\ref{sec:QFT} in frames of QFT, can be
obtained on the basis of Eq.~(\ref{eq:Nut}) as
\begin{align}\label{eq:PeLmuLQM}
  P_{\nu_{e\mathrm{L}}\to\nu_{\mu\mathrm{L}}}(t) = & |\nu_{\mu\mathrm{L}}|^{2}=\frac{\sin^{2}(2\theta)}{4}
  \bigg[
    \cos^{2}(\mu_{1}Bt)+\cos^{2}(\mu_{2}Bt)
    \notag
    \\
    & -
    2\cos(\mu_{1}Bt)\cos(\mu_{2}Bt)\cos\left(\frac{\Delta m^{2}}{2E_{\nu}}t\right)
  \bigg].
\end{align}
To get Eq.~(\ref{eq:PeLmuLQM}) we expand the energies of mass eigenstates
in the ultrarelativistic limit and assume that $E_{\nu}$ is the mean
energy of a neutrino beam.

Now, if we take into account that the distance traveled by a beam
of ultrarelativistic neutrinos is $L\approx t$, as well as identifying
$E_{\nu}$ with $E=(E_{e}+E_{\mu})/2$, we get that the QM transition
probability in Eq.~(\ref{eq:PeLmuLQM}) coincides with the QFT result
in Eq.~(\ref{eq:PnueLmuL}).

\section{Conclusion}\label{sec:CONCL}

In the present work, we have studied the spin-flavor precession of
Dirac neutrinos having diagonal magnetic moments in frames of the
QFT. In the QFT approach in question, which was proposed
in Ref.~\cite{Kob82} originally, neutrinos are virtual particles. Thus, the
contribution of a neutrino to the matrix element results from its
propagator, which is defined for a mass eigenstate only. In case of
neutrinos interacting with a magnetic field owing to the presence
of diagonal magnetic moments, the neutrino propagators should account
for this magnetic interaction. The corresponding propagator was obtained
in Ref.~\cite{Dvo26d}. Note that, in the situation when magnetic
moments are diagonal, the propagators of mass eigenstates are also
diagonal with respect to the particle types. The case of nondiagonal
propagators takes place when a transition magnetic moment is present.
The description of the spin-flavor pressesion in this case was made
in Ref.~\cite{Dvo26b} for Majorana neutrinos and in Ref.~\cite{Dvo26a}
for Dirac ones.

In Sec.~\ref{sec:QFT}, we have derived the matrix element $\mathcal{M}_{e\to\mu}$
corresponding to the channel $\nu_{e\mathrm{L}}\to\nu_{\mu\mathrm{L}}$,
which the transition probability depends on. This matrix element has
been obtained assuming that the charged leptons are the plane waves corresponding
to ultrarelativistic particles. This assumption allowed us to avoid
the momentum nonconservation problem arising in this formalism. Then,
the assumption of the ultrarelativity of neutrinos enabled us to get
rid of the antiparticle contribution to the matrix element. The justification
of this procedure is given in Ref.~\cite{Dvo26c}.

We have also considered rather large distance between a neutrino source
and a detector. In this case, we have obtained only the leading contribution
$\propto L^{-2}$ to the transition probability. The next to leading
QFT corrections to the transition probability, which are sizable at
short macroscopic distances, can be also derived for this channel of
neutrino oscillations. However, is it the subject of a separate study.
Analogous QFT terms were studied, e.g., in Ref.~\cite{NauShk21}
(see also Ref.~\cite{Dvo26a}) to explain the reactor antineutrino
anomaly.

Eventually, in Sec.~\ref{sec:QM}, we have derived the probability
for the channel $\nu_{e\mathrm{L}}\to\nu_{\mu\mathrm{L}}$ in frames
of the QM; cf. Eq.~(\ref{eq:PeLmuLQM}). It has been demonstrated
that the QM and the QFT, given in Eq.~(\ref{eq:PnueLmuL}), expressions
for the transition probability coincide for ultrarelativistic particles.

\end{document}